\documentclass[aps,twocolumn,preprintnumbers]{revtex4}

\usepackage{graphicx}  
\usepackage{epstopdf}
\usepackage{subfigure}
\usepackage{multirow}

\usepackage{fancyhdr}
\usepackage{longtable}
\usepackage{parskip}
\usepackage[T1]{fontenc}
\usepackage{dcolumn}   

\usepackage{bm}        
\usepackage{amsfonts}  
\usepackage{amsmath}   
\usepackage{amssymb}   
\usepackage{siunitx}
\usepackage{tabularx}
\usepackage{setspace}
\usepackage{indentfirst}
\usepackage{float}

\newcommand{\pwisein}{\left\{ \begin{array}{ll}}
\newcommand{\pwiseout}{\end{array}\right.}

\begin{document}

\title{Nuclear fission in intense laser fields}

\author{Jintao Qi,\textsuperscript{1} Libin Fu,\textsuperscript{1} and Xu Wang\textsuperscript{1,}}

\email[\textsuperscript{}]{xwang@gscaep.ac.cn}

\affiliation {\textsuperscript{1}Graduate School, China Academy of Engineering Physics, Beijing 100193, China}

\date{today}

\begin{abstract}
Rapid-advancing intense laser technologies enable the possibility of a direct laser-nucleus coupling. In this paper the effect of intense laser fields on a series of nuclear fission processes, including proton decay, $\alpha$ decay, and cluster decay, is theoretically studied with the help of nuclear double folding potentials. The results show that the half-lives of these decay processes can be modified by non-negligible amounts, for example on the order of 0.01 or 0.1 percents in intense laser fields available in the forthcoming years. In addition to numerical results, an approximate analytical formula is derived to connect the laser-induced modification to the decay half-life and the decay energy. 
\end{abstract}

\maketitle

\section{Introduction}

The intriguing possibility of using intense laser fields to influence and steer nuclear processes has attracted attentions recently \cite{Keitel2013, Misicu2013, Delion2017, Palffy2019, Qi2019, Queisser2019, Lv2019, Wang2020}. This possibility is driven by rapid advancement of intense laser technologies for the past several decades, especially since the invention of the chirped pulse amplification technique \cite{Strickland1985}. Laser fields with peak intensities on the order of 10$^{22}$ W/cm$^2$ can be generated nowadays, and further enhancements for another one or two orders of magnitude are expected in the near future, e.g. with the extreme light infrastructure (ELI) of Europe \cite{Ur2015,Bala2017,Bala2019} or with the superintense ultrafast laser facility (SULF) of Shanghai \cite{Li2018, Yu2018, Zhang2020}.

Although light, especially that with frequencies in the $\gamma$ ray regime, is an important component in traditional nuclear physics, laser has not overlapped much with nuclear physics. Even in strong-laser-initiated inertial confinement fusion \cite{Betti2016}, the connection between laser and nuclear physics is rather indirect, in that laser is just employed to compress the deuterium-tritium fuel cell to a high-density-high-temperature state with which nuclear fusion is more likely to happen. This is understandable because typical nuclear energy levels are on the order of 1 MeV, whereas the energy of a laser photon is on the order of 1 eV, six orders of magnitude smaller.

The gap between laser and nuclear physics, however, starts to be filled as laser is getting more and more intense. When the intensity reaches the order of 10$^{23}$ to 10$^{24}$ W/cm$^2$, laser light can influence nuclear physics in a direct manner. Two characteristic physical quantities may be used to support this assessment. First, the electric field strength corresponding to 10$^{24}$ W/cm$^2$ is comparable to the Coulomb field strength from a nucleus at a distance of about 100 fm away. Intense laser fields therefore push the frontier close to the nuclear territory. Second, the ponderomotive energy (cycle-averaged kinetic energy) of a proton in an 800-nm laser field of intensity 10$^{23}$ W/cm$^2$ is over 3 MeV, reaching energy magnitudes of nuclear physics. 

Indeed, recent works have considered possible influences of intense laser fields on nuclear processes such as $\alpha$ decay \cite{Keitel2013, Misicu2013, Delion2017, Palffy2019, Qi2019} and deuteron-triton fusion \cite{Queisser2019, Lv2019, Wang2020}. These results show that intense laser fields can have non-negligible, in some cases even substantial, {\it direct} influences on these nuclear processes. 

The goal of the current paper is to study possible influences of intense laser fields on some nuclear fission processes, including proton decay \cite{Delion200601, May2011}, $\alpha$ decay, and cluster decay \cite{Poenaru1985, Saidi2015}. These processes can be understood similarly using a quantum tunneling picture following Gamow's treatment of $\alpha$ decay \cite{Gamow1928}. Since tunneling is very sensitive to the potential between the emitted nucleus and the remaining nucleus, and the external laser field modifies this potential, it is not unnatural to expect that there will be some effect on the decay processes. 

To calculate this effect quantitatively, we first construct effective potentials between the emitted nucleus and the remaining nucleus using highly accurate double folding potentials. Then we explain how to include the effect of an external laser field, as well as how to calculate laser-induced modifications to the decay half-life. The numerical results show that with an intensity of 10$^{24}$ W/cm$^2$, the half-lives of these decay processes can be modified by amounts on the order of 0.01 to 0.1 percents. Besides the numerical results, we derive an approximate analytical formula to connect the laser-induced modification to the decay half-life and the decay energy.

This paper is organized as follows. In Sec. \uppercase\expandafter{\romannumeral2} we introduce the methods that we use in our calculations. These include the construction of effective nucleus-nucleus potentials, the inclusion of laser-nucleus interaction, and the calculation of the half-lives. Numerical results on laser-induced modifications to the half-lives, analytical understandings, and discussions will be given in Sec. \uppercase\expandafter{\romannumeral3}. A conclusion will be given in Sec. \uppercase\expandafter{\romannumeral4}.

\section{Method}

\subsection{Effective nucleus-nucleus potentials}

In the past decades, various potential models have been proposed to study nuclear fission processes, e.g., potentials in the Woods-Saxon form \cite{John1958,Buck1992,Buck1994,Talou1999,Hasan2013}, generalized liquid drop models \cite{Royer1985,Royer2000,Zhang2008}, double folding potentials \cite{Petrovich1975,Petrovich1977,Chaudhuri198601,Chaudhuri198602,Abele1993,Hoyler1994,Basu2003}, etc. In this paper, we use the double folding potentials to describe the effective potential between the emitted nucleus and the remaining nucleus.

\subsubsection{The double folding potentials}

The short-range nuclear potential between the emitted nucleus (subscript 1) and the remaining nucleus (subscript 2) is calculated using the following double integral
\begin{equation}
  V_{N}(r)=\int\int\rho_{1}(r_{1})\rho_{2}(r_{2})t_{D}(E,s,\rho_{1},\rho_{2})d^{3}r_{1}d^{3}r_{2}. \label{e.VNdouble}
\end{equation}
Here $\vec{r}_1$ ($\vec{r}_2$) is a vector originated from the center of mass (CM) of nucleus 1 (2), $s=\left| \vec{r}+\vec{r}_{2}-\vec{r}_{1}\right|$ is the distance between the two nuclear mass elements located at $\vec{r}_1$ and $\vec{r}_2$, with $\vec{r}$ the vector from the CM of nucleus 1 to that of nucleus 2. $\rho_{1}(r_{1})$ and $\rho_{2}(r_{2})$ are the mass density distributions of nucleus 1 and 2, respectively, in the following form \cite{Basu2003}
\begin{equation}
\rho_{i}(r_{i})=\frac{\rho_{A}}{1+\exp(\frac{r_i-c_i}{a}) },
\end{equation}
where $\rho_A = 0.165$ fm$^{-3}$, $a=0.54$ fm, and $c_i = 1.13 A_i^{1/3} (1-\pi^2 a^2/3.83A_i^{2/3})$ fm with $A_i$ the mass number of the $i$th nucleus.

The term $t_{D}(E,s,\rho_{1},\rho_{2})$ in Eq. (\ref{e.VNdouble}) is given by \cite{Kobos1984}
\begin{equation}\label{e.td}
  t_{D}(E,s,\rho_{1},\rho_{2})=g(E,s)f_{D}(E,\rho_{1},\rho_{2}),
\end{equation}
where $E$ is the kinetic energy of the emitted nucleus, which is related to the decay energy $Q$ by $E=Q A_{2}/(A_{1}+A_{2})$. The factor $g(E,s)$ is an effective nucleus-nucleus interaction given by \cite{Bertsch1977}
\begin{equation}
  g(E,s)= \left[ 7999\dfrac{e^{-4.0s}}{4.0s}-2134\dfrac{e^{-2.5s}}{2.5s}+J_{00}(E)\delta(s) \right] \text{MeV},
\end{equation}
where $J_{00}(E)$ is a zero-range pseudopotential representing the single-nucleon exchange effect in the form
\begin{equation}
  J_{00}(E)=-276(1-0.005E/A_{1}) \ \text{MeV\,fm$^3$}.
\end{equation}
The mass density-dependent term $f_{D}(E,\rho_{1},\rho_{2})$ in Eq. (\ref{e.td}) is given by \cite{Khoa1997}
\begin{equation}
f_{D}(E,\rho_{1},\rho_{2})=C\left[1+\alpha e^{-\beta(\rho_{1}+\rho_{2})}\right],
\end{equation}
with $C=0.2963$, $\alpha=3.7231$, and $\beta=3.7384$.

The coulomb potential $V_{C}(r)$ is calculated via another double integral
\begin{equation}
V_{C}(r)=\int\int\rho_{1}^{'}(r_{1})\rho_{2}^{'}(r_{2}) \frac{e^2}{s} d^{3}r_{1}d^{3}r_{2},
\end{equation}
where $\rho_{1}^{'}$ and $\rho_{1}^{'}$ are the charge density distributions of the emitted nucleus and the remaining nucleus, respectively. They are given in the following form
\begin{equation}
\rho_{i}^{'}(r_{i})=\frac{\rho_{Z}^{i}}{1+\exp(\frac{r_i-c_i}{a}) }.
\end{equation}
The parameters $a$ and $c_i$ are the same as given above for mass distributions. $\rho_Z^i = 0.165 $ fm$^{-3} \, Z_i/A_i$ is a charge density parameter.

\begin{figure*}[htb]
	\includegraphics[width=0.32\textwidth,totalheight=0.25\textwidth]{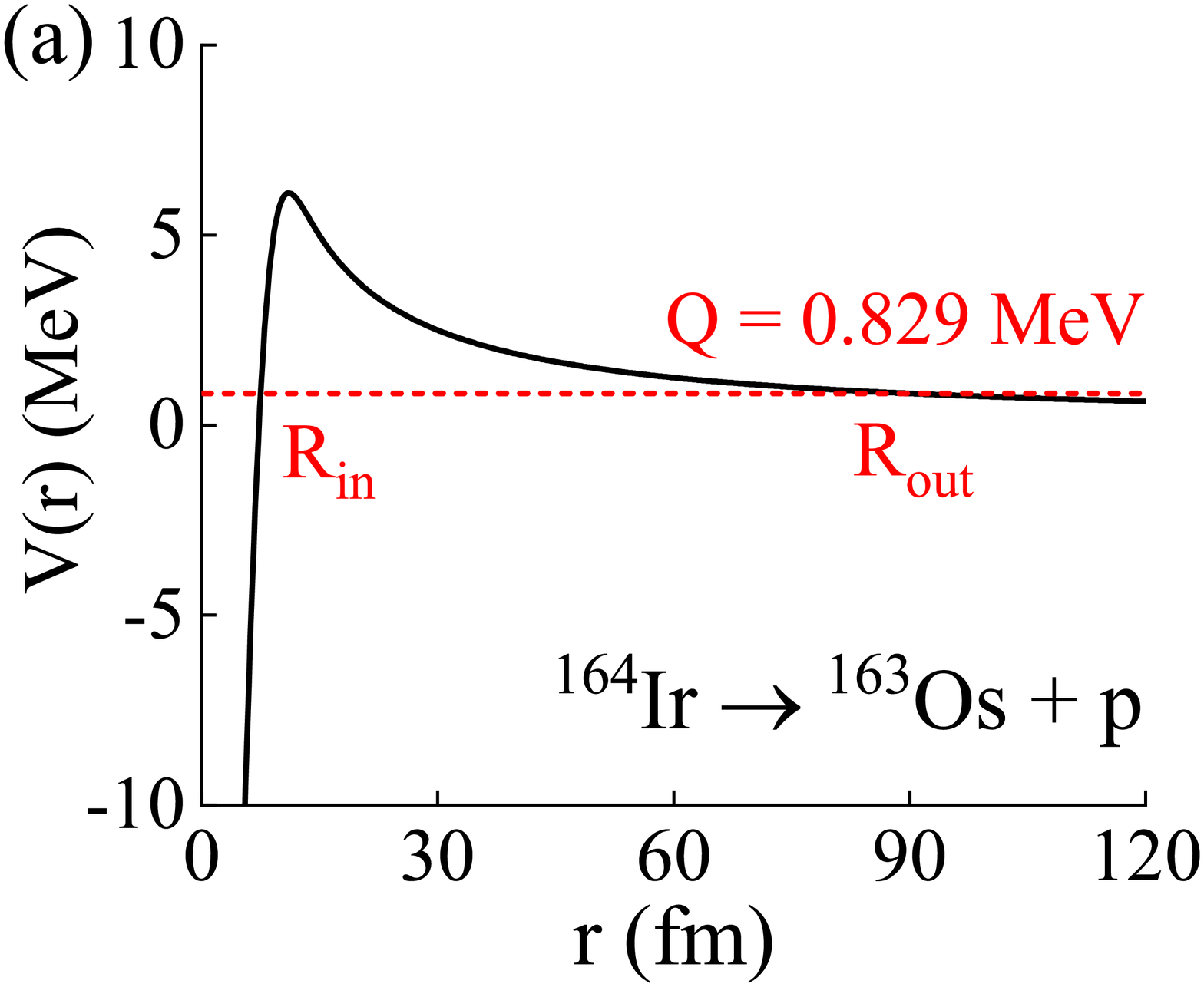}
	\includegraphics[width=0.32\textwidth,totalheight=0.25\textwidth]{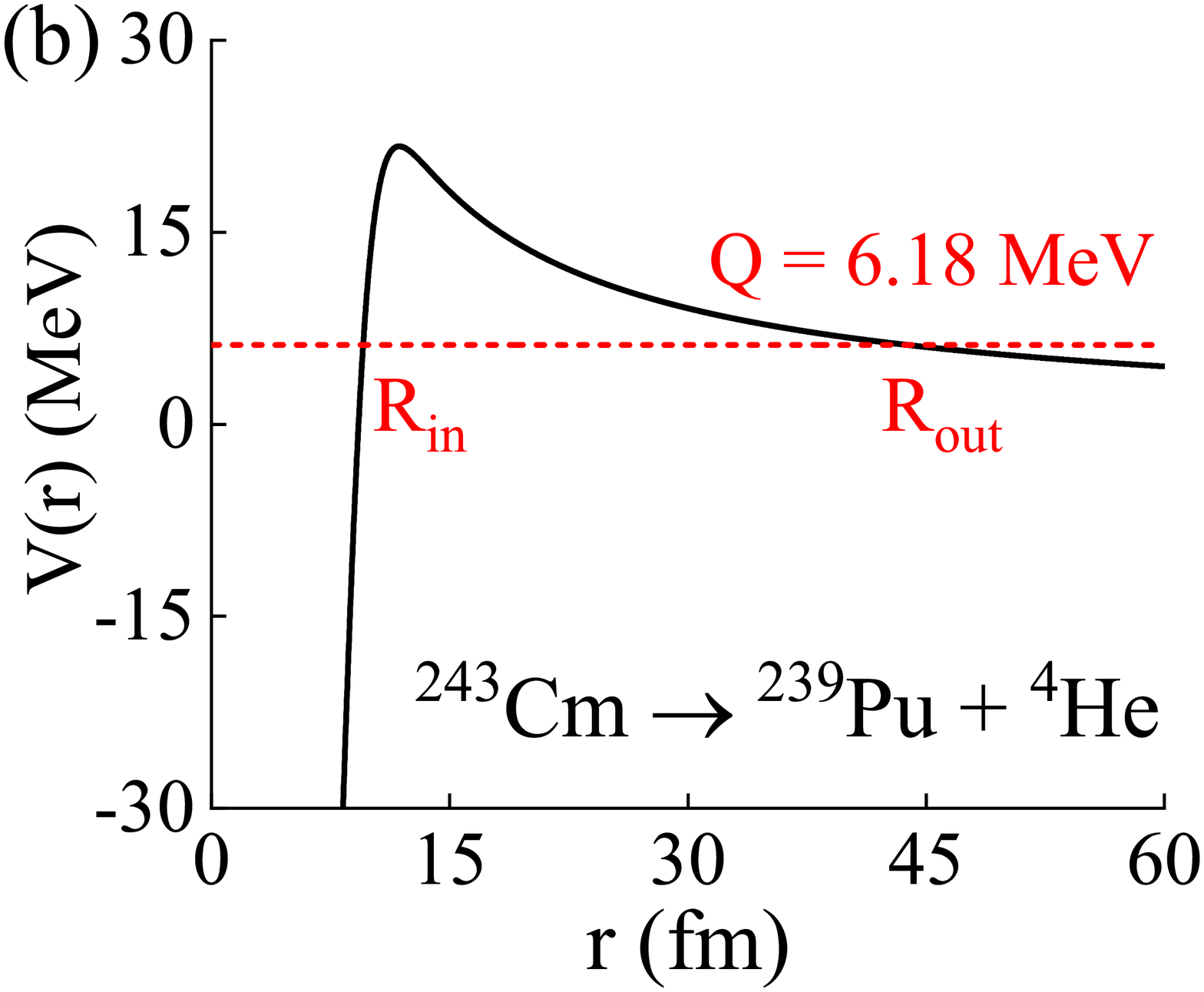}
	\includegraphics[width=0.32\textwidth,totalheight=0.25\textwidth]{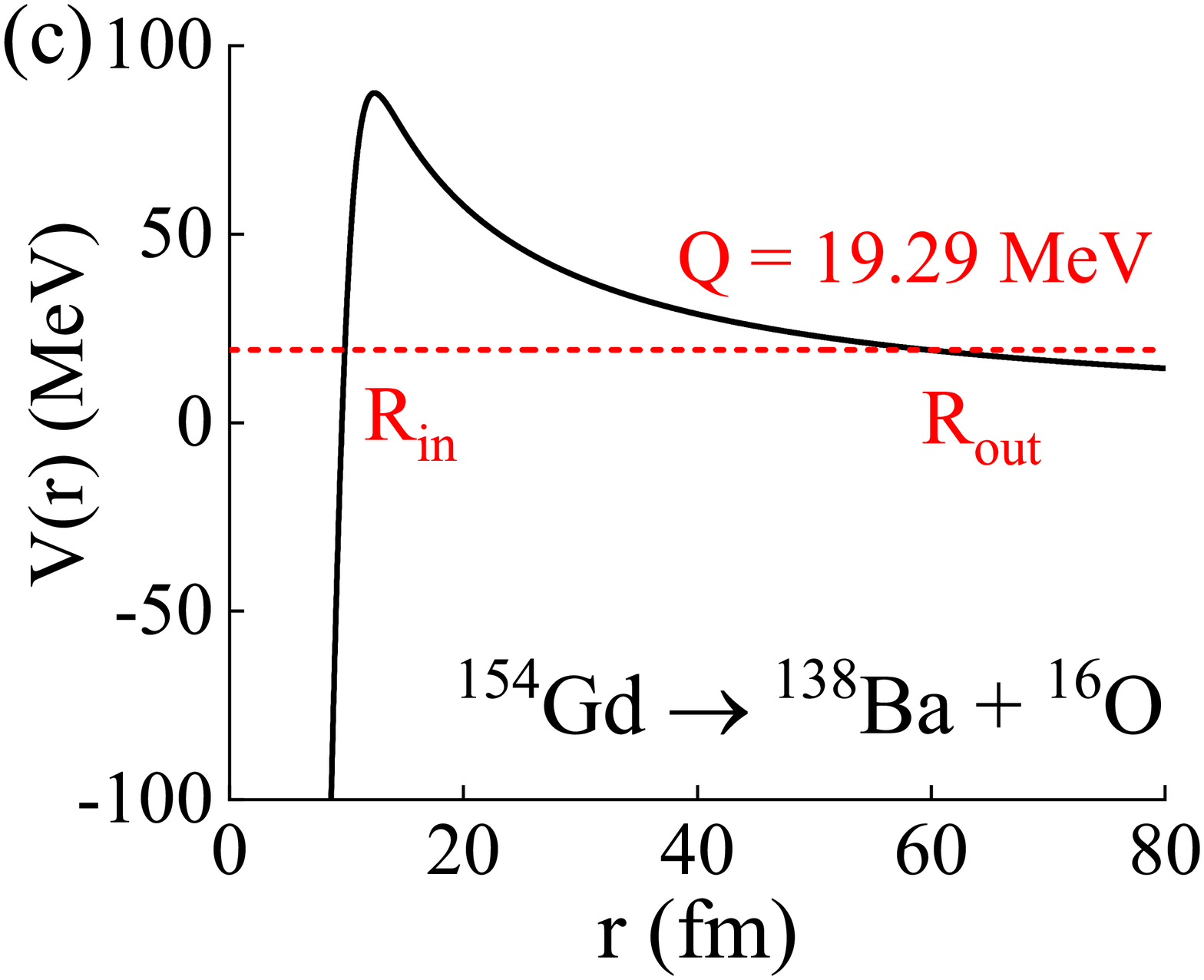}
	\caption{Potential between the emitted nucleus and the remaining nucleus for (a) an example proton decay, (b) an example $\alpha$ decay, and (c) an example cluster decay. The horizontal dashed line in each figure shows the height of the corresponding decay energy $Q$. $R_{in}$ and $R_{out}$ are the tunneling entrance point and exit point determined by the condition $V(r)=Q$. }	\label{f.Vr}
\end{figure*}

For proton decay, the proton can be treated as a point charge. Then the double integral in Eq. (\ref{e.VNdouble}) is simplified to a single integral
\begin{equation}
  V_{N}(r)=\int\rho_{2}(r_{2})t_{S}(E,s,\rho_{2})d^{3}r_{2},
\end{equation}
with $s=|\vec{r}+\vec{r}_2|$. The term $t_{S}(E,s,\rho_{2})$ just involves the density of the remaining nucleus  $ t_{S}(E,s,\rho_{2})=g(E,s)f_{S}(E,\rho_{2})$, where $f_{S}(E,\rho_{2})=C\left[1+\alpha e^{-\beta\rho_{2}}\right]$. The parameters $C$, $\alpha$ and $\beta$ are the same as given above.
Similarly the Coulomb potential between the proton and the remaining nucleus becomes
\begin{equation}
V_{C}(r)=\int\rho_{2}^{'}(r_{2}) \frac{e^2}{s} d^{3}r_{2}.
\end{equation}

\subsubsection{The total nucleus-nucleus potential}

The total potential between the emitted nucleus and the remaining nucleus is
\begin{equation}
  V(r)=V_{N}(r)+V_{C}(r)+\frac{l(l+1)\hbar^{2}}{2\mu r^{2}}, \label{e.Vr}
\end{equation}
where the last term is a centrifugal potential due to the angular momentum of the emitted nucleus, and $\mu=m_{1}m_{2}/(m_{1}+m_{2})$ denotes the reduced mass.

Figure \ref{f.Vr} shows examples of $V(r)$ for (a) proton decay $^{164}$Ir $\rightarrow$ $^{163}$Os + p, (b) $\alpha$ decay $^{243}$Cm $\rightarrow$ $^{239}$Pu + $^4$He, and (c) $^{16}$O-cluster decay $^{154}$Gd $\rightarrow$ $^{138}$Ba + $^{16}$O. The corresponding angular momentum quantum numbers are $l=$ 5, 2, and 0, respectively. The decay energy for each case is shown on figure. From the picture of quantum tunneling, the tunneling entrance point $R_{in}$ and the tunneling exit point $R_{out}$ can be determined using the cross points between the $V(r)$ curve and the horizontal line of $Q$. One sees that the length of tunneling ($R_{out}$ - $R_{in}$) is usually several tens of fm.

\subsection{Laser-nucleus interaction}

\subsubsection{The center-of-mass coordinates}

The two-nucleus system in a laser field can be described by a time-dependent Schr\"odinger equation (TDSE)
\begin{equation}
  i\hbar\frac{\partial\psi\left(\vec{r}_{1},\vec{r}_{2},t\right)}{\partial t}=H\left(t\right)\psi\left(\vec{r}_{1},\vec{r}_{2},t\right),
\end{equation}
where the minimum-coupling Hamiltonian is
\begin{equation}
\begin{split}
  H(t)=
  & \sum\limits_{i=1,2} \frac{1}{2m_{i}}\left[\vec{p}_{i}-q_{i}\vec{A}(t)\right]^{2}+V\left(r \right).
\end{split}
\end{equation}
A radiation gauge has been used with which the scalar potential of the laser field vanishes. We have also applied a dipole approximation with which the spatial dependency of the vector potential can be neglected. This is justified by the fact that the spatial scale of concern to nuclear fission processes is on the order of 10 to 100 fm (see Fig. \ref{f.Vr}), which is much smaller than the laser wavelength. Existing strong laser fields may be in the near-infrared regime with wavelengths around 800 nm, or in the X-ray regime of wavelengths on the order of 1 nm from free-electron lasers. 

For fission processes it is more convenient to switch to the CM coordinates ($\vec{R}, \vec{r}, \vec{P}, \vec{p}$):
\begin{align}
    \vec{r}_{1}&=\vec{R}+\frac{m_{2}}{m_{1}+m_{2}}\vec{r}, &\vec{r}_{2}&=\vec{R}-\frac{m_{1}}{m_{1}+m_{2}}\vec{r}, \\
    \vec{p}_{1}&=\vec{p}+\frac{m_{1}}{m_{1}+m_{2}}\vec{P}, &\vec{p}_{2}&=-\vec{p}+\frac{m_{2}}{m_{1}+m_{2}}\vec{P}.
\end{align}
Then the Hamiltonian can be written as
\begin{equation}
H(t) = \frac{1}{2M}\left[\vec{P}-q\vec{A}(t)\right]^{2}+\frac{1}{2\mu}\left[\vec{p}-q_{eff}\vec{A}(t)\right]^{2}+V\left(r \right),
\end{equation}
where $q=q_1+q_2$, $M=m_1+m_2$, $\mu$ is the reduced mass, and $q_{eff} = (q_1m_2-q_2m_1)/(m_1+m_2)$ is an effective charge for relative motion. $V(r)$ is not affected.

The wave function can also be expressed in the CM coordinates $\psi(\vec{r}, \vec{R},t)$. By introducing unitary transformations
\begin{equation}
  \varphi(\vec{r}, \vec{R}, t)=\hat{U_{r}}\hat{U_{R}}\psi(\vec{r}, \vec{R}, t),
\end{equation}
where $\hat{U}_r=\exp[-iq_{eff}\vec{A}(t)\cdot\vec{r}/\hbar]$ and $\hat{U}_R=\exp[-iq\vec{A}(t)\cdot\vec{R}/\hbar]$, the TDSE can be written as
\begin{equation}
\begin{split}
i\hbar\frac{\partial\varphi(\vec{r},\vec{R},t)}{\partial t}= &
\left[
-\frac{\hbar^2}{2\mu}\nabla_r^2+V(r)-q_{eff}\vec{r}\cdot\vec{\varepsilon}(t) \right.\\
&\left.-\frac{\hbar^2}{2M}\nabla_R^2-q\vec{R}\cdot\vec{\varepsilon}(t)
\right]
\varphi(\vec{r},\vec{R},t).
\end{split}
\end{equation}
Here $\vec{\varepsilon}(t) = -d\vec{A}(t)/dt$ is the laser electric field.

A factorization of the wavefunction $\varphi(\vec{r},\vec{R},t)=\phi(\vec{r},t)\chi(\vec{R},t)$ can be performed, and the TDSE can be separated into two independent equations: one for the CM
\begin{equation}
i\hbar\frac{\partial\chi(\vec{R},t)}{\partial t}=\left[ -\frac{\hbar^2}{2M}\nabla_{R}^{2}-q\vec{R}\cdot\vec{\varepsilon}(t) \right]\chi(\vec{R},t),
\end{equation}
and the other for the relative motion
\begin{equation}
i\hbar\frac{\partial\phi(\vec{r},t)}{\partial t}=\left[ -\frac{\hbar^2}{2\mu}\nabla_{r}^{2}+V(r)-q_{eff}\vec{r}\cdot\vec{\varepsilon}(t) \right]\phi(\vec{r},t).
\end{equation}
It is the latter equation that will be of relevance to the fission processes considered here. The interaction potential energy between the laser field and the relative motion particle is
\begin{equation}
V_I(r, \varepsilon, \theta) = -q_{eff} \vec{r} \cdot \vec{\varepsilon}(t) = -q_{eff} r \varepsilon(t) \cos\theta, \label{e.VI}
\end{equation}
where we have assumed that the laser field is linearly polarized along the $z$ axis. $\theta$ is the angle between the emission direction and the $+z$ axis.

\subsubsection{The quasistatic condition}

For nuclear fission, a preformation picture is commonly used which assumes that the later-emitted nucleus has been preformed inside its parent nucleus before emission. From typical decay energies (around 1 MeV for proton decay, several MeV for $\alpha$ decay, tens to hundreds MeV for cluster decay) one can estimate that the velocity of a preformed nucleus is on the order of 10$^{7}$ m/s, a small fraction of the speed of light. The size of the parent nucleus is about 1 fm. In the preformation picture, the later-emitted nucleus oscillates back and forth inside its parent nucleus. The frequency of this oscillation can be roughly estimated to be 10$^{7}$ m s$^{-1}$/1 fm $=$ 10$^{22}$ Hz. Every time it hits the potential wall, the preformed nucleus has a very small probability of tunneling out. 

The length of the tunneling path is usually between 10 and 100 fm. The time for the emitted nucleus to travel through the tunneling region can be estimated to be $100$\,fm$/10^7$\,m s$^{-1}= 10^{-20}$ s. This time is much smaller than an optical period of currently available strong lasers. For near-infrared lasers, the wavelength is around 800 nm and a laser cycle is about $10^{-15}$ s. For X-ray free electron lasers with photon energy 10 keV, a laser cycle is on the order of $10^{-19}$ s. Therefore, during the time for the emitted nucleus to travel through the potential barrier, the laser field does not have time to change appreciably and can be viewed as quasistatic. This quasistatic condition has also been discussed previously by us in Ref. \cite{Qi2019}. We mention that the quasistatic condition has been widely used in strong-field atomic physics to describe tunneling ionization of atoms \cite{Ammosov1986,Corkum1996,Chen2000}.

\begin{table*}[!htbp]
	
	\centering
	\begin{spacing}{1.5}
		\caption{Laser-induced modifications to the half-lives of selected fission processes, including proton decay, $\alpha$ decay, and cluster decay. The laser field is assumed to have an intensity of 10$^{24}$ W/cm$^{2}$. The decay half-lives without the laser field are obtained from Refs. \cite{Delion2006, Qian2014, Basu2003, Poenaru1985}. } \label{t.p-a-h}
		\begin{tabularx}{17cm}{p{1.4cm}<{\raggedright} p{1.4cm}<{\centering} p{1.4cm}<{\centering} p{1.5cm}<{\centering} p{2.3cm}<{\centering} p{2.8cm}<{\centering} p{2.8cm}<{\centering} p{2.3cm}<{\centering}}
			\hline\hline
			Parent & Emitted & $l$ & $Q$ (MeV) & $P_{0}$ & $T$($\varepsilon=0$) (s) & $T$($\varepsilon$, $\theta=0$) (s) & $\Delta_{T}$ \\
		\end{tabularx}
		\begin{tabularx}{17cm}{p{1.4cm}<{\raggedright} p{1.4cm}<{\centering} p{1.4cm}<{\centering} p{1.5cm}<{\centering} p{2.3cm}<{\centering} p{2.8cm}<{\centering} p{2.8cm}<{\centering} p{2.3cm}<{\raggedleft}}
			\hline
			
			$_{53}^{109}$I & p & 2 & 0.83 & 3.095$\times$10$^{-1}$ & 1.03039$\times$10$^{-4}$ & 1.02696$\times$10$^{-4}$ & 3.329$\times$10$^{-3}$ \\
			$_{55}^{113}$Cs & p & 2 & 0.98 & 9.741$\times$10$^{-2}$ & 1.67109$\times$10$^{-5}$ & 1.66711$\times$10$^{-5}$ & 2.387$\times$10$^{-3}$ \\
			$_{67}^{141}$Ho & p & 3 & 1.20 & 5.417$\times$10$^{-2}$ & 4.10204$\times$10$^{-3}$ & 4.09276$\times$10$^{-3}$ & 2.263$\times$10$^{-3}$ \\
			$_{71}^{150}$Lu & p & 5 & 1.28 & 5.352$\times$10$^{-1}$ & 6.60693$\times$10$^{-2}$ & 6.59214$\times$10$^{-2}$ & 2.239$\times$10$^{-3}$ \\
			$_{73}^{155}$Ta & p & 5 & 1.79 & 4.367$\times$10$^{-1}$ & 1.19950$\times$10$^{-5}$ & 1.19822$\times$10$^{-5}$ & 1.066$\times$10$^{-3}$ \\

			$_{60}^{144}$Nd & $_{2}^{4}$He & 0 & 1.90 & 1.722$\times$10$^{-1}$ & 6.60693$\times$10$^{+22}$ & 6.58914$\times$10$^{+22}$ & 2.693$\times$10$^{-3}$  \\
			$_{72}^{164}$Hf & $_{2}^{4}$He & 0 & 5.28 & 3.114$\times$10$^{-2}$ & 2.45471$\times$10$^{+2}$ & 2.45426$\times$10$^{+2}$ & 1.847$\times$10$^{-4}$  \\
			$_{88}^{221}$Ra & $_{2}^{4}$He & 2 & 6.89 & 2.338$\times$10$^{-2}$ & 2.81838$\times$10$^{+1}$ & 2.81760$\times$10$^{+1}$ & 2.790$\times$10$^{-4}$  \\
			$_{90}^{217}$Th & $_{2}^{4}$He & 5 & 9.43 & 1.649$\times$10$^{-3}$ & 2.51189$\times$10$^{-4}$ & 2.51160$\times$10$^{-4}$ & 1.127$\times$10$^{-4}$  \\
			$_{90}^{222}$Th & $_{2}^{4}$He & 0 & 8.13 & 3.207$\times$10$^{-2}$ & 2.81838$\times$10$^{-3}$ & 2.81788$\times$10$^{-3}$ & 1.799$\times$10$^{-4}$  \\
			$_{90}^{232}$Th & $_{2}^{4}$He & 0 & 4.08 & 1.549$\times$10$^{-1}$ & 4.46684$\times$10$^{+17}$ & 4.46152$\times$10$^{+17}$ & 1.192$\times$10$^{-3}$  \\
			$_{96}^{243}$Cm & $_{2}^{4}$He & 2 & 6.18 & 2.225$\times$10$^{-3}$ & 9.12011$\times$10$^{+8}$ & 9.11587$\times$10$^{+8}$ & 4.646$\times$10$^{-4}$  \\

			$_{62}^{150}$Sm & $_{6}^{12}$C & 0 & 11.21 & 1.551$\times$10$^{-11}$ & 6.30957$\times$10$^{+48}$ & 6.30105$\times$10$^{+48}$ & 1.352$\times$10$^{-3}$ \\
			$_{64}^{154}$Gd & $_{8}^{16}$O & 0  & 19.29 & 9.014$\times$10$^{-12}$ & 3.16228$\times$10$^{+48}$ & 3.15936$\times$10$^{+48}$ & 9.236$\times$10$^{-4}$ \\
			$_{88}^{223}$Ra & $_{6}^{14}$C & 4 & 31.85 & 5.125$\times$10$^{-10}$ & 1.58489$\times$10$^{+15}$ & 1.58473$\times$10$^{+15}$ & 1.008$\times$10$^{-4}$ \\
			$_{90}^{226}$Th & $_{8}^{18}$O & 0 & 45.73 & 7.362$\times$10$^{-12}$ & 6.30957$\times$10$^{+16}$ & 6.30865$\times$10$^{+16}$ & 1.462$\times$10$^{-4}$ \\
			$_{92}^{235}$Th & $_{10}^{24}$Ne & 1 & 57.36 & 1.204$\times$10$^{-14}$ & 2.51189$\times$10$^{+27}$ & 2.51163$\times$10$^{+27}$ & 1.039$\times$10$^{-4}$ \\
			$_{92}^{236}$U & $_{12}^{28}$Mg & 0 & 71.83 & 3.317$\times$10$^{-17}$ & 3.98107$\times$10$^{+27}$ & 3.98045$\times$10$^{+27}$ & 1.553$\times$10$^{-4}$ \\
			$_{94}^{236}$Pu & $_{12}^{28}$Mg & 0 & 79.67 & 2.884$\times$10$^{-18}$ & 5.01187$\times$10$^{+21}$ & 5.01139$\times$10$^{+21}$ & 9.666$\times$10$^{-5}$ \\	
			$_{94}^{238}$Pu & $_{14}^{32}$Si & 0 & 91.21 & 9.820$\times$10$^{-20}$ & 1.99526$\times$10$^{+25}$ & 1.99495$\times$10$^{+25}$ & 1.551$\times$10$^{-4}$ \\
			\hline
		\end{tabularx}
	\end{spacing}
\end{table*}

\subsection{Penetrability and half-life}

In the preformation picture, the later-emitted nucleus oscillates back and forth within its parent nucleus. Every time it hits the potential wall, it has a probability of tunneling out. This probability is called the penetrability. From the quasistatic condition, the penetrability can be calculated for each time (i.e. for each laser field strength $\varepsilon$) using the well-known Wentzel-Kramers-Brillouin (WKB) formula as
\begin{equation}
  P(\varepsilon, \theta)=\exp\left(-2\int_{R_{in}}^{R_{out}}k(r, \varepsilon, \theta)dr\right). \label{e.Penetrability}
\end{equation}
The wavenumber $k(r, \varepsilon, \theta)$ is defined as
\begin{equation}
	k(r, \varepsilon, \theta)=\frac{1}{\hbar} \sqrt{2\mu\left[V(r)-Q+V_{I}(r, \varepsilon, \theta)\right]},
\end{equation}
where $V(r)$ and $V_{I}(r, \varepsilon, \theta)$ have been given above in Eq. (\ref{e.Vr}) and (\ref{e.VI}), respectively.

The decay width of the emitted nucleus is given by \cite{Gurvitz1987}
\begin{equation} \label{e.decaywidth}
  \Gamma(\varepsilon, \theta) = \dfrac{\hbar^{2}}{4\mu} P_{0} F(\varepsilon, \theta) P(\varepsilon, \theta),
\end{equation}
where $P_{0}$ is called the preformation or spectroscopic factor. The so-called normalization factor $F(\varepsilon, \theta)$ is given by
\begin{equation}
  F(\varepsilon, \theta) = \left[ \int_{0}^{R_{in}}\frac{dr}{2 k(r, \varepsilon, \theta)} \right]^{-1},
\end{equation}
which is very insensitive to the external laser field because the integration is performed inside the nucleus from 0 to $R_{in}$. One can safely treat $F$ as a laser-independent constant. With the decay width, the half-life is given as
\begin{equation} \label{e.halflife}
T(\varepsilon, \theta) = \hbar \ln2/\Gamma(\varepsilon, \theta).
\end{equation}

The preformation factor $P_0$ has been studied extensively \cite{Berg1997,Mohr2006,Ren2010} and is usually given to gap the difference between the calculated half-life and the experimentally measured value. That is, $P_0$ is set to a value such that $T(\varepsilon=0) = T^{exp}(\varepsilon=0)$. With $\varepsilon=0$ all directions are the same, so we may omit the $\theta$ argument.

In this paper, we use the relative change of the half-life, noted $\Delta_{T}$, to describe the effect of an intense laser field. $\Delta_{T}$ is defined as
\begin{equation} \label{e.DeltaT}
  \Delta_{T}= \left| \frac{T(\varepsilon, \theta=0) - T(\varepsilon=0)}{T(\varepsilon=0)} \right|. 
\end{equation}
With the presence of the laser field $\varepsilon$, we will focus on the forward direction $\theta=0$. Other directions are equivalent to smaller laser field strengths of $\varepsilon \cos\theta$ (and in the opposite direction if $\theta>90^{\circ}$). In the forward direction the half-life will be a little bit smaller than the laser-free half-life. Nevertheless for the sake of simplicity, we take the absolute value on the right hand side of Eq. (\ref{e.DeltaT}) such that $\Delta_T$ is always positive.

\section{results and discussions}

\subsection{Laser-induced modifications to half-lives}

TABLE \ref{t.p-a-h} shows numerical results of laser-induced modifications to half-lives, $\Delta_T$, of selected proton decay, $\alpha$ decay, and cluster decay processes. The decay mode, angular momentum, decay energy, and preformation factor have been listed for clarity. The laser intensity is assumed to be 10$^{24}$ W/cm$^{2}$, which is expected to be achievable in the forthcoming years. One sees that $\Delta_{T}$ (rightmost column) is on the order of 0.1\% for proton decay, and mostly on the order of 0.01\% to 0.1\% for $\alpha$ decay and cluster decay. 

\begin{figure}[tbp]
	\centering
	\includegraphics[scale=0.3]{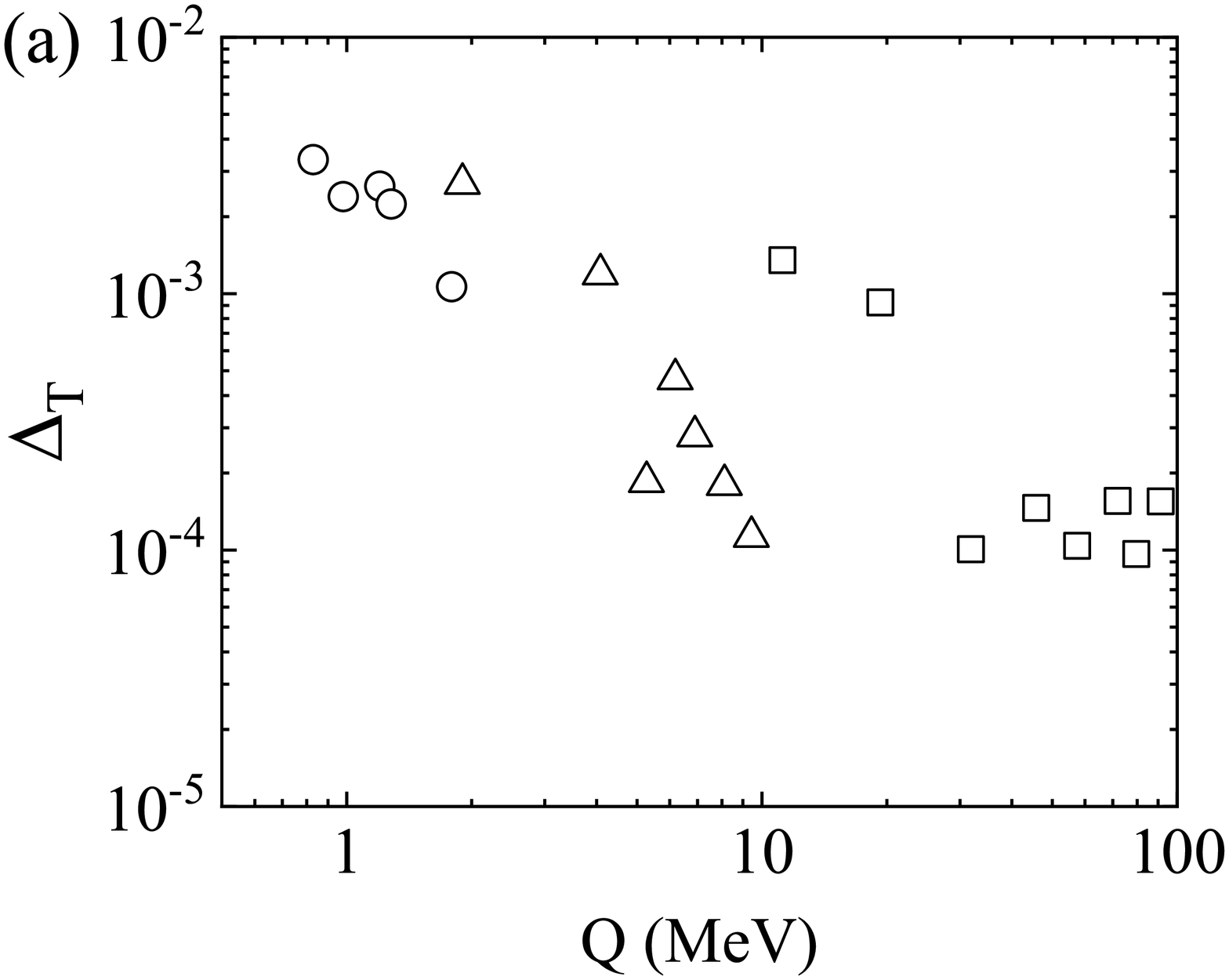}
	\includegraphics[scale=0.3]{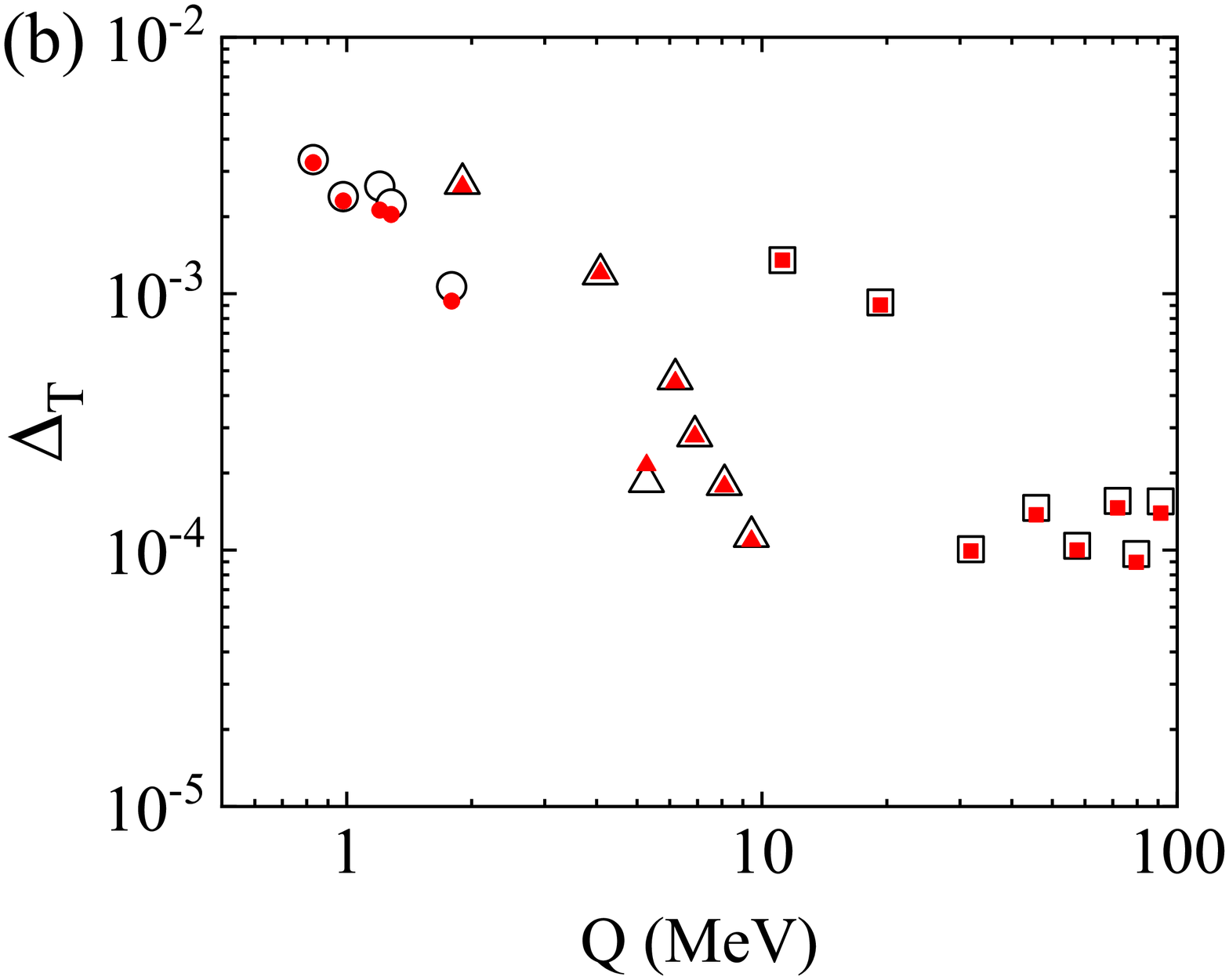}
	\caption{ (a) Relative change of half-lives, $\Delta_T$, of proton decay (circles), $\alpha$ decay (triangles), and cluster decay (squares), as a function of the decay energy $Q$. Both axes are in the logarithmic scale. (b) Approximate analytical predictions from Eq. (\ref{e.GNlaw22}) are shown as (red) filled symbols. The positions of the unfilled symbols are the same as in (a).} \label{f.modifications}
\end{figure}

Figure \ref{f.modifications} (a) shows, graphically, $\Delta_T$ as a function of the decay energy $Q$ for proton decay (circles), $\alpha$ decay (triangles), and cluster decay (squares). One sees more clearly from this figure that proton decays are, in general, easier to be modified by external laser fields. The reason is that proton decays usually have lower Coulomb barriers and longer tunneling paths (from $R_{in}$ to $R_{out}$) for the laser field to act on. The general trend is that the larger the decay energy $Q$, the smaller the laser-induced modification $\Delta_T$. However, other factors, such as the decay type, are also important. For example, $\alpha$ decay and cluster decay with similar decay energies (around 10 MeV) can have very (over an order of magnitude) different values of $\Delta_T$.

\subsection{An approximate formula connecting $\Delta_T$ and $Q$}

Next we try to provide analytical insights into the numerical results given in TABLE \ref{t.p-a-h} and Fig. \ref{f.modifications} (a), by using simplified nucleus-nucleus potentials such that analytical treatments are possible. We start from Eq. (\ref{e.Penetrability}) for the penetrability, and treat the laser interaction $V_I (r, \varepsilon, \theta)$ as a perturbation to the remaining potential $V_0(r) \equiv V(r) - Q$. Then the penetrability
\begin{align}
P(\varepsilon, \theta) &=\exp\left(-\frac{2\sqrt{2\mu}}{\hbar}\int_{R_{in}}^{R_{out}}\sqrt{V_{0}}\sqrt{1+\frac{V_{I}}{V_{0}}}dr\right) \nonumber\\
 &\approx \exp \left[ -\frac{2\sqrt{2\mu}}{\hbar}\int_{R_{in}}^{R_{out}}\sqrt{V_{0}} \left( 1+\frac{V_{I}}{2V_{0}} \right) dr \right] \nonumber\\
&= \exp( \gamma^{(0)}+\gamma^{(1)} ).
\end{align}
A Taylor expansion has been performed from the first step to the second step, and $\gamma^{(0)}$ and $\gamma^{(1)}$ are shorthand notations defined as
\begin{align}
\gamma^{(0)}&=-\frac{2\sqrt{2\mu}}{\hbar}\int_{R_{in}}^{R_{out}}\sqrt{V_{0}}dr,   \label{e.gamma0}\\
\gamma^{(1)}&=\varepsilon\frac{\sqrt{2\mu}q_{eff}\cos\theta}{\hbar}\int_{R_{in}}^{R_{out}}\frac{rdr}{\sqrt{V_{0}}}. \label{e.gamma1}
\end{align}
In the latter formula the expression of $V_I$ in Eq. (\ref{e.VI}) has been used. 

In general Eqs. (\ref{e.gamma0}) and (\ref{e.gamma1}) can only be integrated numerically. To continue we consider a simplified version of the nuclear potential $V(r)$ instead of the elaborated one given in Eq. (\ref{e.Vr}) that has been used above for numerical calculations. We consider $V(r)$ to be a square potential well for $r<R_{in}$ plus a Coulomb potential $q_{1}q_{2}/r$ for $r>R_{in}$. Here $R_{in}=1.13(A_{1}^{1/3}+A_{2}^{1/3})$ fm is the geometrical touching distance. The depth of the square potential well will not be of concern if only the penetrability is to be calculated. The tunneling exit point can be determined as $R_{out}=q_{1}q_{2}/Q$. Besides, the centrifugal potential is ignored in this simplified version of $V(r)$. Then $\gamma^{(0)}$ and $\gamma^{(1)}$ can be integrated analytically. For the former
\begin{align}
\gamma^{(0)}&=-\frac{2\sqrt{2\mu}}{\hbar} Q^{1/2} \int_{R_{in}}^{R_{out}}\sqrt{V(r)/Q-1}dr \nonumber\\
&=-\frac{2\sqrt{2\mu}q_{1}q_{2}}{\hbar} Q^{-1/2} \left[ \eta-\frac{1}{2}\sin (2\eta)\right] \nonumber\\
&\approx -a Q^{-1/2} - b
\end{align}
where $\eta \equiv \cos^{-1} \sqrt{R_{in}/R_{out}}$, and a Puiseux series expansion has been performed for the terms in the square bracket. The coefficients $a$ and $b$ are given as
\begin{align}
a= \pi\sqrt{2\mu}q_{1}q_{2} / \hbar,\ \ \ \ \ \
b= - 4\sqrt{2\mu q_{1}q_{2}R_{in}}/\hbar.
\end{align}
Using Eqs. (\ref{e.decaywidth}) and (\ref{e.halflife}) one can get
\begin{align}
\ln T = a Q^{-1/2} + b',   \label{e.GNlaw}
\end{align}
where $b' = b + \ln (4 \mu \ln2 / \hbar P_0 F)$. This is the famous Geiger-Nuttall law \cite{Geiger1911} that connects the decay half-life $T$ and the decay energy $Q$.

\begin{figure}[tbp]
	\centering
	\includegraphics[scale=0.3]{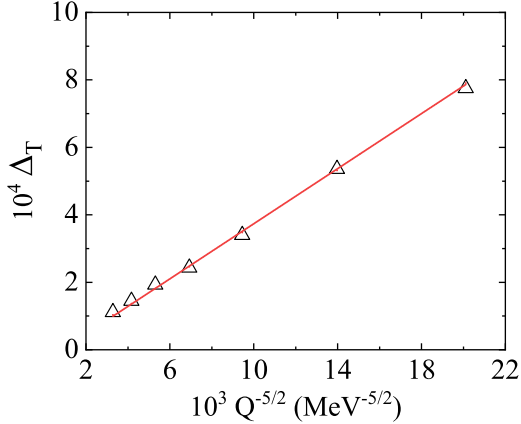}
	\caption{ $\Delta_T$ vs. $Q^{-5/2}$ for $\alpha$ decay of even-even $^{218-230}$Th isotopes. The unfilled symbols are numerical results, and they fall approximately on a line (The line is added just to guide the eye).} \label{f.linear}
\end{figure}

Next we consider $\gamma^{(1)}$:
\begin{align}
\gamma^{(1)}&=\varepsilon\frac{\sqrt{2\mu}q_{eff}\cos\theta}{\hbar} Q^{-1/2} \int_{R_{in}}^{R_{out}}\frac{rdr}{\sqrt{V(r)/Q-1}} \nonumber\\
&=\varepsilon\frac{2\sqrt{2\mu}q_{eff}q_{1}^{2}q_{2}^{2}\cos\theta}{\hbar}Q^{-5/2} \nonumber\\
&\quad\times\left[ \frac{3}{8}\eta+\frac{1}{4}\sin(2\eta)+\frac{1}{32}\sin(4\eta)\right] \nonumber\\
&\approx cQ^{-5/2} + d.
\end{align}
Again, a Puiseux series expansion has been performed from the second step to the third step for the terms in the square bracket. The coefficients $c$ and $d$ are given as
\begin{align}
c&= \varepsilon\frac{3\pi\sqrt{2\mu}q_{eff}q_{1}^{2}q_{2}^{2}\cos\theta}{8\hbar},   \label{e.parac}\\ 
d&= -\varepsilon\frac{4\sqrt{2\mu}q_{eff}R_{in}^{5/2}\cos\theta}{5\hbar\sqrt{q_{1}q_{2}}}. \label{e.parad}
\end{align}
Using the relation $\Delta_T = 1 - P(\varepsilon=0)/P(\varepsilon)$, we get
\begin{align}
\ln(1-\Delta_T) = - cQ^{-5/2} - d. \label{e.GNlaw2}
\end{align}
Noticing $\Delta_T \ll 1$, one may Taylor expand the left hand side and get 
\begin{align}
\Delta_T = cQ^{-5/2} + d. \label{e.GNlaw22}
\end{align}

The $\Delta_T$ values obtained using Eq. (\ref{e.GNlaw22}) are shown in Fig. \ref{f.modifications}(b) as (red) filled symbols, to be compared with the numerical results. In general the agreements are fairly good, confirming the validity of the above approximations and analyses. A few cases do have visible discrepancies with the numerical results due to the usage of the simplified nucleus-nucleus potential and the neglect of the centrifugal potential.

For decays from isotopes of an element, e.g. $\alpha$ decay from \{$^{218}$Th, $^{220}$Th, $^{222}$Th, $^{224}$Th, $^{226}$Th, $^{228}$Th, $^{230}$Th\}, the coefficients $\{c, d\}$ are rather close [See Eqs. (\ref{e.parac}-\ref{e.parad})]. One may expect for these decays that the \{$\Delta_T$, $Q$\} pairs fall (approximately) on a line on a $\Delta_T$-versus-$Q^{-5/2}$ plot. This is indeed the case, as shown in Fig. \ref{f.linear}.

\section{conclusions}

To summarize, we report a theoretical and numerical study of effects of intense laser fields on a series of nuclear fission processes, including proton decay, $\alpha$ decay, and cluster decay. We provide a complete theoretical framework, including the construction of effective nucleus-nucleus potentials, the inclusion of the laser-nucleus interaction, and the calculation of laser-induced modifications to decay half-lives. Fission processes that are not included in the current study can easily be calculated using this framework, if needed.

We arrive at the conclusion that with intense laser fields to be expected in the forthcoming years, e.g. with an intensity of 10$^{24}$ W/cm$^{2}$, the nuclear fission processes can be modified by small, yet, finite amounts on the order of 0.01 to 0.1 percents. These amounts seem not big, but certainly cannot be simply ignored. {\it The important message to deliver is that laser starts to be able to directly influence nuclear physics}. This is remarkable especially considering that the energy of a single laser photon is so small, and the ability of the influence is provided by the extremely high achievable intensities. There is certainly much to be expected if laser becomes an efficient tool to influence and eventually to control nuclear physics.

{\bf Acknowledgments}. This work was supported by Science Challenge Project of China No. TZ2018005, NSFC No. 11774323, and NSAF No. U1930403.

\end{document}